\documentclass[showpacs,floatfix,superscriptaddress,nofootinbib,aps,prl,twocolumn]{revtex4}
\usepackage[dvips]{graphicx}
\usepackage{latexsym,amssymb,amsmath,epsfig,bm,times,psfrag,subfigure}
\usepackage{color}
\usepackage[bookmarksnumbered,bookmarksopen,colorlinks,citecolor=red,linkcolor=blue]{hyperref}
\renewcommand\a{\alpha}
\renewcommand\b{\beta}

\renewcommand\r{\rho}

\newcommand\g{\gamma}

\newcommand\p{\pi}
\newcommand\h{\theta}

\newcommand\f{\phi}

\newcommand\D{\Delta}

\newcommand\J{\Psi}

\newcommand{\fig}[1]{Fig.~\ref{#1}}

\newcommand\lb{\left(}
\newcommand\rb{\right)}
\newcommand\ls{\left[}
\newcommand\rs{\right]}

\newcommand{\lan}{\langle}
\newcommand{\ran}{\rangle}

\newcommand{\non}{\nonumber\\}

\newcommand{\rp}{{\rm RP}}

\renewcommand{\part}{{\rm part}}
\newcommand{\calp}{{\mathcal P}}
\newcommand{\calc}{{\mathcal C}}

\newcommand{\be}{\begin{equation}}
\newcommand{\ee}{\end{equation}}
\newcommand{\bear}{\begin{eqnarray}}
\newcommand{\eear}{\end{eqnarray}}
\newcommand{\ba}{\begin{array}}
\newcommand{\ea}{\end{array}}

\begin{document}

\title{Test the chiral magnetic effect with isobaric collisions}
\author{Wei-Tian Deng}
\affiliation{School of physics, Huazhong University of Science and Technology, Wuhan 430074, China}
\author{Xu-Guang Huang}
\affiliation{Physics Department and Center for Particle Physics and Field Theory, Fudan University, Shanghai 200433, China}
\affiliation{Department of Physics, Brookhaven National Laboratory, Upton, New York 11973-5000, USA}
\author{Guo-Liang Ma}
\affiliation{Shanghai Institute of Applied Physics, Chinese Academy of Sciences, Shanghai 201800, China}
\author{Gang Wang}
\affiliation{Department of Physics and Astronomy, University of California, Los Angeles, California 90095, USA}


\begin{abstract}
The quark-gluon matter produced in relativistic heavy-ion collisions may contain local domains in which $\calp$ and $\calc\calp$
symmetries are not preserved. When coupled with an external magnetic field, such $\calp$- and $\calc\calp$-odd domains will
generate electric currents along the magnetic field --- a phenomenon called the chiral magnetic effect (CME).
Recently, the STAR Collaboration at RHIC and the ALICE Collaboration at the LHC released data of charge-dependent
azimuthal-angle correlators with features consistent with the CME expectation. However, the experimental observable
is contaminated with significant background contributions from elliptic-flow-driven effects, which makes the interpretation
of the data ambiguous. In this Letter, we show that the collisions of isobaric nuclei, $^{96}_{44}$Ru + $^{96}_{44}$Ru
and $^{96}_{40}$Zr + $^{96}_{40}$Zr, provide an ideal tool to disentangle the CME signal from the background effects.
Our simulation demonstrates that the two collision types at $\sqrt{s_{\rm NN}}=200$ GeV have more than $10\%$ difference in the CME signal
and less than $2\%$ difference in the elliptic-flow-driven backgrounds for the centrality range of $20-60\%$.
\end{abstract}
\maketitle

Quantum chromodynamics (QCD), the modern theory of the strong interaction, permits the violation of parity symmetry ($\calp$) or
combined charge conjugation and parity symmetry ($\calc\calp$), although accurate experiments performed so far have not seen
such violation at vanishing temperature and density~\cite{Baker:2006ts}. Recently it was suggested that in the hot and dense
matter created in high-energy heavy-ion collisions, there may exist metastable domains where $\calp$ and $\calc\calp$ are
violated owing to vacuum transitions induced by topologically nontrivial gluon fields, e.g., sphalerons~\cite{Kharzeev:1998kz}.
In such a domain, net quark chirality can emerge from chiral anomaly, and the strong magnetic field of a non-central collision
can then induce an electric current along the magnetic field, which is known as the chiral magnetic effect
(CME)~\cite{Kharzeev:2007jp,Fukushima:2008xe}; see Refs.~\cite{Huang:2015oca,Kharzeev:2015znc} for recent reviews of the
magnetic field and the CME in heavy-ion collisions.

The CME provides a means to monitoring the topological sector of QCD, and the experimental search for the CME has been intensively
performed in heavy-ion collisions at RHIC and the LHC. To detect the CME, a three-point correlator,
\begin{eqnarray}
\label{corre}
\g_{\a\b}=\lan\cos(\f_\a+\f_\b-2\J_\rp)\ran,
\end{eqnarray}
was proposed~\cite{Voloshin:2004vk}, where $\f$ is the azimuthal angle of a charged particle,
the subscript $\a$ ($\b$) denotes the charge sign of the particle (positive or negative),
$\J_\rp$ is the angle of the reaction plane of a given event,
and $\lan\cdots\ran$ denotes an average over all particle pairs and all the events.
The occurrence of the CME driven by the magnetic field (perpendicular to the reaction plane) is expected to contribute
a positive opposite-sign (OS) correlator and a negative same-sign (SS) correlator. The measurements of the correlator $\g$
by the STAR Collaboration for Au + Au collisions at $\sqrt{s_{\rm NN}}=200$ GeV~\cite{Abelev:2009ac,Abelev:2009ad} and
by the ALICE Collaboration for Pb + Pb collisions at $\sqrt{s_{\rm NN}}=2.76$ TeV~\cite{Abelev:2012pa},
indeed demonstrate the expected features of the CME.
The signal is robust against various ways of determination of the reaction plane, and persists when the collision system
changes to Cu + Cu or U + U, and when the collision energy is lowered down to
$\sqrt{s_{\rm NN}}=19.6$ GeV~\cite{Abelev:2009ad,Adamczyk:2013hsi,Wang:2012qs,Adamczyk:2014mzf}.
For further lowered collision energies, the difference between
$\g_{\rm OS}$ and $\g_{SS}$ steeply falls down~\cite{Adamczyk:2014mzf}, which may be understood by noticing that at
lower energies the system is probably in a hadronic phase where the chiral symmetry is broken and the CME is strongly suppressed.

Ambiguities, however, exist in the interpretation of the experimental results, owing to possible background effects
that are not related to the CME, e.g., local charge conservation~\cite{Pratt:2010gy,Schlichting:2010na,Schlichting:2010qia},
transverse momentum conservation~\cite{Pratt:2010gy,Pratt:2010zn,Bzdak:2010fd}, etc. These background effects,
once coupled with elliptic flow ($v_2$)~\cite{Poskanzer}, will contribute to $\g_{\a\b}$. To disentangle the possible CME signal and the
flow-related backgrounds, one can utilize experimental setups to either vary the backgrounds with the signal fixed,
or vary the signal with the backgrounds fixed.

The former approach was carried out by exploiting the prolate shape of the uranium nuclei~\cite{Voloshin:2010ut}.
In central U + U collisions, one expects sizable $v_2$ but a negligible magnetic field,
and thus a vanishingly small CME contribution to the correlator $\g$. The STAR Collaboration collected $0-1\%$ most
central events from U + U collisions at $\sqrt{s_{\rm NN}}=197$ GeV in 2012, and indeed found sizable $v_2$ while
the difference between $\g_{\rm OS}$ and $\g_{\rm SS}$ (note that the charge-blind backgrounds are subtracted in $\D\g$),
\begin{eqnarray}
\label{delt_corre}
\D\g\equiv\g_{\rm OS}-\g_{\rm SS},
\end{eqnarray}
is consistent with zero~\cite{Wang:2012qs}. However, it was found that the total multiplicity of detected hadrons is far less
dependent on the number of binary collisions than expected~\cite{Adamczyk:2015obl}, so it is very hard to isolate tip-tip collisions
(that generate small $v_2$) from body-body collisions (that generate large $v_2$). This significantly reduces the lever arm
available to manipulate $v_2$ in order to separate $v_2$-driven backgrounds from the CME.

\begin{figure}[!htb]
\begin{center}
\includegraphics[height=3.5cm]{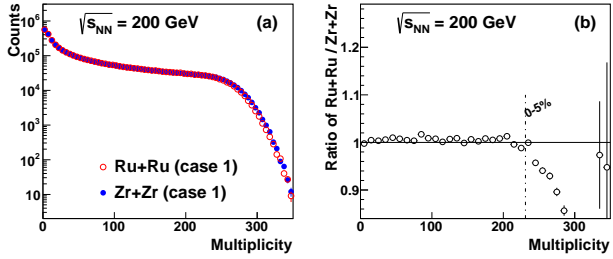}
\caption{The Monte Carlo Glauber simulation of the multiplicity distributions for $^{96}_{44}$Ru + $^{96}_{44}$Ru and $^{96}_{40}$Zr + $^{96}_{40}$Zr at $\sqrt{s_{\rm NN}}=200$ GeV (a) and their ratio (b).}
\label{fig_multi}
\end{center}
\end{figure}

The latter approach (with the $v_2$-driven backgrounds fixed) can be realized, especially for mid-central/mid-peripheral events,
with collisions of isobaric nuclei, such as $^{96}_{44}$Ru and $^{96}_{40}$Zr~\cite{Voloshin:2010ut}.
Ru + Ru and Zr + Zr collisions at the same beam energy are almost identical in terms of particle production,
which is illustrated with the Monte Carlo Glauber simulation~\cite{Adcox:2000sp,Bearden:2001xw,Back:2002uc,Adams:2003yh}
in \fig{fig_multi}. The ratio of the multiplicity distributions from the two collision systems is consistent with unity
almost everywhere, except in $0-5\%$ most central collisions, where the slightly larger radius of Ru ($R_0 = 5.085$ fm)
plays a role against the smaller radius of Zr ($R_0 = 5.02$ fm). Our centrality bins are defined
with the same multiplicity cuts for the two collision types. For the CME analysis, we focus on the centrality range
of $20-60\%$, so that the background contributions due to the multiplicity is negligible.

The spatial distribution of nucleons in either $^{96}_{44}$Ru or $^{96}_{40}$Zr in the rest frame can be written
in the Woods-Saxon form (in spherical coordinates),
\begin{eqnarray}
\r(r,\h)=\frac{\r_0}{1+\exp{[(r-R_0-\b_2 R_0 Y^0_2(\h))/a]}},
\end{eqnarray}
where $\r_0=0.16$ fm$^{-3}$ is the normal nuclear density, $R_0$ and $a$ represent the ``radius" of the nucleus and
the surface diffuseness parameter, respectively, and $\b_2$ is the deformity of the nucleus. The parameter $a$ is almost identical
for Ru and Zr: $a\approx 0.46$ fm. Our current knowledge of $\b_2$ of Ru and Zr is incomplete. There are two sources of available
information on $\b_2$: e-A scattering experiments~\cite{Raman:1201zz,Pritychenko:2013gwa} and comprehensive model
deductions~\cite{Moller:1993ed}. According to the first source (which will be referred to as case 1), Ru is more deformed
($\b_2^{\rm Ru} = 0.158$) than Zr ($\b_2^{\rm Zr} = 0.08$); while the second source (which will be referred to as case 2)
tells the opposite, $\b_2^{\rm Ru}=0.053$ is smaller than $\b_2^{\rm Zr}=0.217$. As we have checked, this systematic uncertainty
has little influence on the multiplicity distribution. We will discuss later its noticeable impacts on the CME signal
(via the magnetic field) and the $v_2$-driven backgrounds (via $\epsilon_2$, the initial spatial eccentricity of the
participant zone).

\begin{figure}[!htb]
\begin{center}
\includegraphics[height=3.5cm]{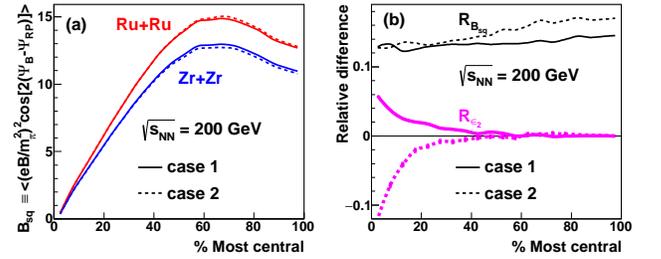}
\caption{Theoretical calculation of the initial magnetic field squared with correction from azimuthal fluctuation for Ru + Ru and Zr + Zr collisions at $\sqrt{s_{\rm NN}}=200$ GeV (a) and their relative difference (b) versus centrality. Also shown is the relative difference in initial eccentricity (b). The solid (dashed) lines correspond to the parameter set of case 1 (case 2).}
\label{fig_mag}
\end{center}
\end{figure}

The charge difference between Ru and Zr nuclei provides a handle on the initial magnetic field (mostly produced by the spectator
protons)~\cite{Skokov:2009qp,Deng:2012pc}. Figure~\ref{fig_mag}(a) presents the theoretical calculation of the initial magnetic
field squared with correction from azimuthal fluctuation of the magnetic field orientation,
$B_{sq}\equiv\lan(eB/m_\p^2)^2\cos[2(\J_{\rm B}-\J_{\rm RP})]\ran$ (with $m_\p$ the pion mass and $\J_{\rm B}$ the azimuthal angle
of the magnetic field), for the two collision systems at 200 GeV, using the HIJING model~\cite{Deng:2012pc,Deng:2014uja}.
$B_{sq}$ quantifies the magnetic field's capability of driving the CME signal in the correlator
$\g$~\cite{Bloczynski:2012en,Bloczynski:2013mca}. For the same centrality bin, the Ru + Ru collision produces a significantly
stronger magnetic field than Zr + Zr. Some theoretical uncertainties come from the modeling of the nuclei, e.g., how to model
the electric charge distribution of the proton: treating the proton as a point charge or as a uniformly charged ball. For the
event averaged calculation, this type of uncertainty is small. Another uncertainty involves the Lienard-Wiechert potential
used in this calculation, which applied no quantum corrections. At RHIC energies, including corrections from quantum
electrodynamics makes little difference~\cite{Huang:2015oca}. The theoretical uncertainties are greatly suppressed
when we take the ratio or relative difference between the two systems. Panel (b) of Fig.~\ref{fig_mag} shows that
the relative difference in $B_{sq}$ between Ru + Ru and Zr + Zr collisions is approaching $15\%$ (case 1) or $18\%$ (case 2)
for peripheral events, and reduces to about $13\%$ (case 1 and case 2) for central
events~\footnote{In our notation, the relative difference in a quantity $F$ between Ru + Ru and Zr + Zr collisions
is $R_F \equiv 2(F^{\rm Ru+Ru}-F^{\rm Zr+Zr})/(F^{\rm Ru+Ru}+F^{\rm Zr+Zr})$, and $F$ can be $B_{sq}$, $\epsilon_2$ or $S$.}. The
effect of the deformity of the nucleus on the generation of the magnetic field is more distinctive in more peripheral collisions.

In \fig{fig_mag}(b), we also show the relative difference in the initial eccentricity, $R_{\epsilon_2}$, obtained from the
Monte Carlo Glauber simulation. $R_{\epsilon_2}$ is highly consistent with 0 for peripheral events, and goes above (below) 0
for the parameter set of case 1 (case 2) in central collisions, because the Ru (Zr) nucleus is more deformed.
The relative difference in $v_2$ should closely follow that in eccentricity, so for the centrality range of interest,
$20-60\%$, the $v_2$-related backgrounds should stay almost the same for Ru + Ru and Zr + Zr collisions.
The slightly non-zero effect will be taken into account in the significance estimation for the CME signal projection,
to be discussed later.

Given the initial magnetic fields and eccentricities, we can estimate the relative difference in the charge-separation observable
$S\equiv N_{\rm part}\D\g $ between Ru + Ru and Zr + Zr collisions. Here $N_{\rm part}$ is used to compensate for the dilution
effect, which is expected when there are multiple sources involved in the collision~\cite{Abelev:2009ad,Ma:2011uma}.
The focus of the isobaric collisions is on the lift of degeneracy between Ru + Ru and Zr + Zr,
therefore we express the corresponding $S$ with a two-component perturbative approach to emphasize the relative difference
\begin{eqnarray}
S^{\rm Ru+Ru} &=& {\bar S}\ls(1-bg)\lb 1+\frac{R_{B_{sq}}}{2}\rb + bg\lb 1+\frac{R_{\epsilon_2}}{2}\rb\rs, \non\\
S^{\rm Zr+Zr} &=& {\bar S}\ls(1-bg)\lb 1-\frac{R_{B_{sq}}}{2}\rb + bg\lb 1-\frac{R_{\epsilon_2}}{2}\rb\rs,\non
\label{eq:S}
\end{eqnarray}
where $bg\in [0,1]$ quantifies the background contribution due to elliptic flow and ${\bar S}=(S^{\rm Ru+Ru}+S^{\rm Zr+Zr})/2$.
An advantage of the perturbative approach is that the relative difference in $S$,
\begin{eqnarray}
R_S = (1-bg)R_{B_{sq}}+ bg\cdot R_{\epsilon_2},
\end{eqnarray}
is independent of the detailed implementation of ${\bar S}$.
Without loss of generality, we parametrize ${\bar S}$ based on the STAR measurements of $S^{\rm Au+Au}$
at 200 GeV~\cite{Adamczyk:2013hsi} as a function of $B_{sq}^{\rm Au+Au}$:
${\bar S} = (2.17+2.67{\bar B}_{sq}-0.074{\bar B}_{sq}^2)\times 10^{-3}$,
where ${\bar B}_{sq} = (B_{sq}^{\rm Ru+Ru} + B_{sq}^{\rm Zr+Zr})/2$. It is noteworthy that the data points of $(S, B_{sq})$ for $30-60\%$ Cu+Cu collisions
at 200 GeV~\cite{Abelev:2009ad} also fall onto this curve. Note that ${\bar S}$ is almost a linear function of ${\bar B}_{sq}$ at small ${\bar B}_{sq}$ values,
because the coefficient of the quadratic term is very small.

\begin{figure}[!htb]
\begin{center}
\includegraphics[height=3.5cm]{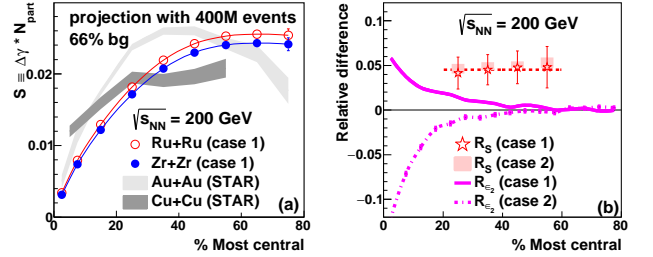}
\caption{Projection of $S\equiv N_{\rm part} \D\g $ for Ru + Ru and Zr + Zr collisions at $\sqrt{s_{\rm NN}}=200$ GeV for the
parameter set of case 1 (a) and the relative difference in the two (b) versus centrality, assuming the background level to be
two thirds. Also shown in panel (b) is the relative difference in the initial eccentricity from the Monte Carlo Glauber simulation
(pink solid and dashed lines). }
\label{fig_gamma}
\end{center}
\end{figure}

In Fig.~\ref{fig_gamma}(a) we show the projection of $S^{\rm Ru+Ru}$ and $S^{\rm Zr+Zr}$ at 200 GeV,
as functions of centrality, with $B_{sq}$ and $\epsilon_2$ obtained for case 1, and the background level $bg=2/3$.
The statistical errors are estimated based on 400 million events for each collision type.
The gray bands depict the STAR measurements of $S^{\rm Au+Au}$ and $S^{\rm Cu+Cu}$ at 200 GeV in comparison.
For $30-60\%$ collisions, all the collision types share a universal curve of $S(B_{sq})$ or ${\bar S}({\bar B}_{sq})$,
which transforms into a rough atomic-number ordering in $S$ as a function of centrality.

The systematic uncertainties in the projection are largely canceled out with the relative difference between Ru + Ru and Zr + Zr,
shown in \fig{fig_gamma}(b); in comparison, we show again the relative difference in eccentricity. For both parameter sets of
the Glauber inputs (red stars for case 1 and pink shaded boxes for case 2), the relative difference in $S$ is about $5\%$
for centrality range of $20-60\%$. The amounts of $R_S$ can be easily guessed from the values of $R_{B_{sq}}$ in \fig{fig_mag}(b)
scaled down by a factor of 3 (since $bg=2/3$ and $R_{\epsilon_2}$ is close to 0).
When we combine the events of $20-60\%$ centralities, $R_S$ is $5\sigma$ above $R_{\epsilon_2}$ for
both parameter sets of the Glauber inputs.
Therefore, the isobaric collisions provide a unique test to pin down the underlying physics mechanism for the observed charge
separation. As a by-product, $v_2$ measurements in central collisions will discern which information source (case 1 or 2) is
more reliable regarding the deformity of the Ru and Zr nuclei.

\begin{figure}[!htb]
\begin{center}
\includegraphics[height=3.9cm]{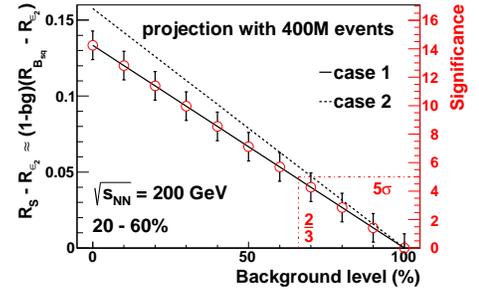}
\caption{Magnitude (left axis) and significance (right axis) of the relative difference in the CME signal between Ru + Ru
and Zr + Zr at 200 GeV, $R_S-R_{\epsilon_2}$ as a function of the background level.}
\label{fig_signi}
\end{center}
\end{figure}

When a different background level is assumed, the magnitude and significance of the projected relative difference between Ru + Ru
and Zr + Zr change accordingly, as shown in \fig{fig_signi}. The measurements of the isobaric collision data will determine
whether there is a finite CME signal observed in the correlator $\g$, and if the answer is ``yes", will ascertain the background
contribution, when compared with this figure. With 400 million events for each collision type, the background level can be determined with an accuracy of $~7\%$. 

In summary, we have numerically simulated the strengths of the initial magnetic fields and the participant eccentricities for
the isobaric collisions of $^{96}_{44}$Ru + $^{96}_{44}$Ru and $^{96}_{40}$Zr + $^{96}_{40}$Zr. Using the previous STAR
measurements of the three-point correlator (\ref{corre}) in Au + Au and Cu + Cu collisions as baseline, we estimate
the relative difference in the charge-separation observable $S=N_{\rm part}\D\g$ between Ru + Ru and Zr + Zr collisions, assuming a background level of two thirds.
We find a noticeable relative difference in $S$ which is robust in the $20-60\%$ centrality bins. Our results strongly suggest
that the isobaric collisions can serve as an ideal tool to disentangle the signal of the chiral magnetic effect from the
$v_2$-driven backgrounds.

Finally, we point out that the isobaric collisions may also be used to disentangle the signal of the chiral magnetic wave
(CMW)~\cite{Kharzeev:2010gd,Burnier:2011bf} from background effects. We summarise in Table \ref{table} the expected relationship
between Ru + Ru and Zr + Zr in terms of experimental observables for elliptic flow, the CME, the CMW and the chiral vortical
effect (CVE)~\cite{Erdmenger:2008rm,Banerjee:2008th,Son:2009tf}, assuming that the chiral effects are the major physical mechanisms
for the corresponding observables.
\begin{table}[h!]
  \centering
  \caption{The expected relationship between Ru + Ru and Zr + Zr in terms of experimental observables for elliptic flow, the CME,
the CMW and the CVE.}
  \label{table}
  \begin{tabular}{|c|c|}
  \hline
    Observable & $^{96}_{44}$Ru + $^{96}_{44}$Ru \;vs.\; $^{96}_{40}$Zr +$^{96}_{40}$Zr\\
    \hline
    flow & \;\;$\approx$ \\
    \hline
    CME & \;\; $>$ \\
    \hline
    CMW & \;\; $>$ \\
    \hline
    CVE & \;\; $\approx$ \\
    \hline
  \end{tabular}
\end{table}

{\bf Acknowledgments:} We are grateful to H.~Huang, D.~Kharzeev, J.~Liao, S.~Voloshin, N.~Xu, and Z.~Xu for helpful
communications and discussions. W.-T.D is supported by NSFC with Grant No. 11405066.
X.-G.H. is supported by NSFC with Grant No. 11535012 and the One Thousand Young Talents Program of China.
G.-L.M is supported by supported by NSFC with Grants No. 11375251, No. 11522547, No. 11421505, and the Major State Basic Research Development Program in China with Grant No. 2014CB845404. G.W is supported by the US Department of Energy under Grants No. DE-FG02-88ER40424.


\end{document}